\documentclass[conference]{IEEEtran}
\usepackage{cite}
\usepackage{graphicx}

\usepackage{hyperref}
\hypersetup{colorlinks,allcolors=black}

\usepackage[table]{xcolor}
\usepackage{caption}
\usepackage{subcaption}
\usepackage{amsmath}
\graphicspath{ {images/} }

\begin{document}


\title{Supervised Learning for Coverage-Directed Test Selection in Simulation-Based Verification}


\author
{\IEEEauthorblockN{Nyasha Masamba}
\IEEEauthorblockA{Faculty of Engineering\\Trustworthy Systems Laboratory\\
University of Bristol\\
United Kingdom\\
Email: nyasha.masamba@bristol.ac.uk }
\and
\IEEEauthorblockN{Kerstin Eder}
\IEEEauthorblockA{Faculty of Engineering\\Trustworthy Systems Laboratory\\
University of Bristol\\
United Kingdom\\
Email: kerstin.eder@bristol.ac.uk }
\and
\IEEEauthorblockN{Tim Blackmore}
\IEEEauthorblockA{Infineon Technologies\\Stoke Gifford\\
Bristol\\
United Kingdom\\
Email: tim.blackmore@infineon.com }
}
\maketitle


\begin{abstract}
\label{abstract}
Constrained random test generation is one of the most widely adopted methods for generating stimuli for simulation-based verification. Randomness leads to test diversity, but tests tend to repeatedly exercise the same design logic. Constraints are written (typically manually) to bias random tests towards interesting, hard-to-reach, and yet-untested logic. However, as verification progresses, most constrained random tests yield little to no effect on functional coverage. If stimuli generation consumes significantly less resources than simulation, then a better approach involves randomly generating a large number of tests, selecting the most effective subset, and only simulating that subset. In this paper, we introduce a novel method for automatic constraint extraction and test selection. This method, which we call coverage-directed test selection, is based on supervised learning from coverage feedback. Our method biases selection towards tests that have a high probability of increasing functional coverage, and prioritises them for simulation. We show how coverage-directed test selection can reduce manual constraint writing, prioritise effective tests, reduce verification resource consumption, and accelerate coverage closure on a large, real-life industrial hardware design.

\end{abstract}

\begin {IEEEkeywords}
Design Verification, Supervised Learning, Coverage-Directed Test Generation, Test Selection, Machine Learning for Verification, CDG, EDA
\end{IEEEkeywords}

\section{Introduction}
\label{introduction}

Functional verification is the process of ensuring the functional correctness of a hardware design with respect to its intended specification \cite{Wileetal2005}. A popular method of performing functional verification is simulation-based verification. Simulation-based verification is a process in which the design under test (DUT) is stressed by test stimuli within a simulation engine, with the resulting DUT behaviour being checked against expected behaviour according to the specification. Deviations between DUT behaviour and the specification indicate possible faults, referred to as `bugs', in the design logic. Exercised DUT functionality is recorded in a coverage database. Coverage is an important measure of verification progress and test quality. Based on analysis of the coverage reports, verification effort can focus on uncovered areas. 

\begin{figure}[t]
    \centering
    \includegraphics[width=0.49\textwidth]{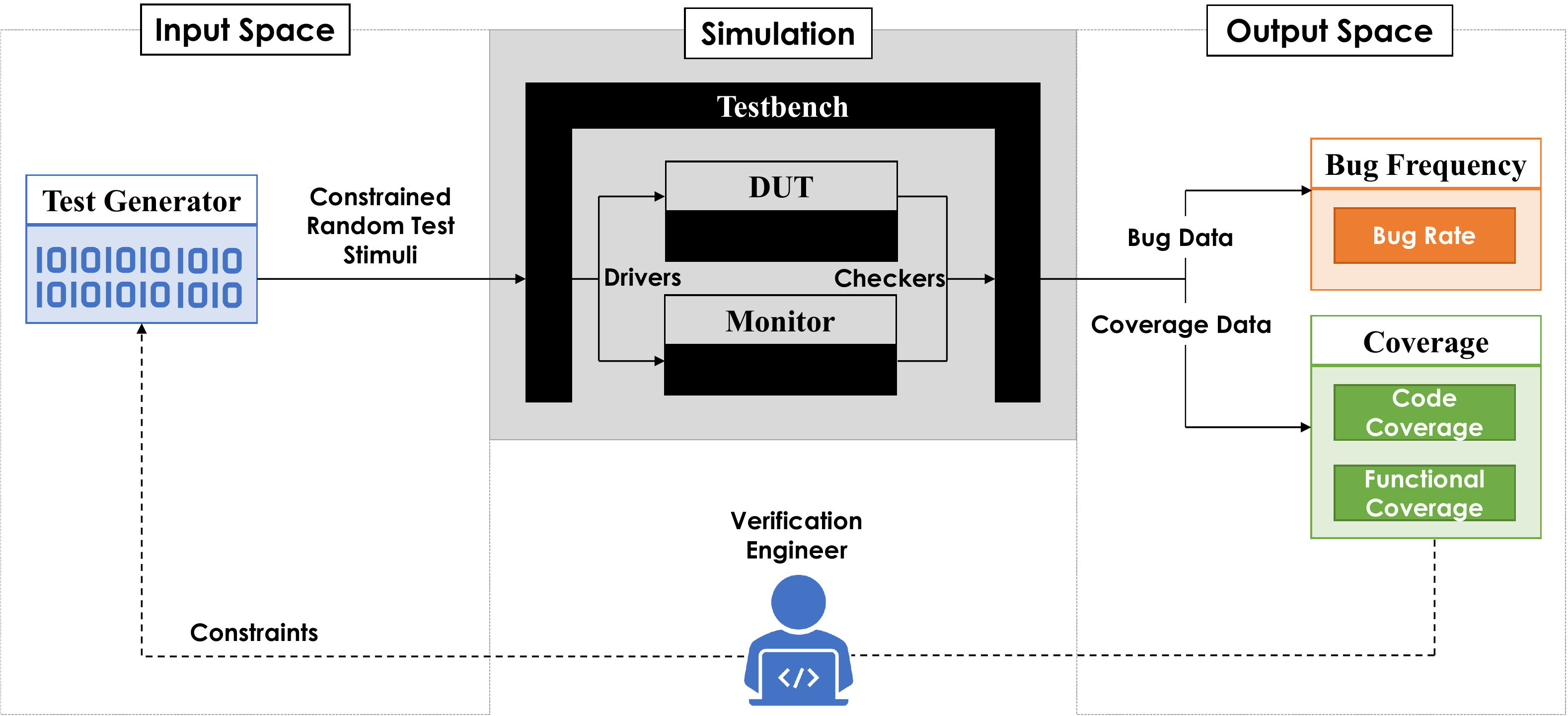}
    \caption{Constrained Random Test Generation}
    \label{fig:traditional_crv}
\end{figure}

Constrained random test generation, illustrated in Figure \ref{fig:traditional_crv}, is currently the state-of-the-art test generation method. Random test generation allows for automatic generation of diverse test stimuli. Constraining random tests ensures they are legal and meaningful with respect to the specification. However, many of the generated tests are redundant in that they repeatedly explore the same DUT functionality while consuming significant verification resources. Constrained random tests are generally ineffective at exercising hard-to-reach or yet uncovered DUT functionality. 

A logical unit of DUT functionality that has been identified for verification and remains unexercised is called a coverage hole. Writing constraints that target coverage holes is tedious work often carried out manually by senior verification engineers with deep knowledge of the DUT on a trial-and-error basis. 

Automatic test biasing based on coverage feedback leads to more effective testing. Coverage feedback consists of coverage data derived from coverage reports, and previously simulated test data.  Effective tests are those that are legal, meaningful, and useful at exercising the DUT functionality required to plug coverage holes. Effective testing significantly increases functional verification productivity by freeing engineers from manually biasing test generation. Coverage is also closed faster, resulting in faster project completion.

In this paper, we introduce coverage-directed test selection (CDS) as a novel, machine learning-based \cite{Flach2012} approach to automatic test biasing during simulation-based verification. Although we largely focus on using the method for test selection, our approach can also be adapted for test generation through techniques such as constraint extraction using decision tree classification. The CDS method is applicable to any design in which labelled coverage feedback data can be presented in a tabular structure (i.e., with features as columns and rows as training examples), and where simulation can be decoupled from coverage collection.

This paper is structured as follows. Section \ref{related_work} introduces related research, and how the limitations of the research motivated this work. Section \ref{cds} provides a detailed introduction to CDS. Section \ref{experiment} describes the environment in which the CDS experiments were conducted. Section \ref{results} presents and discusses the results derived from performing CDS using a variety of supervised learning algorithms. Section \ref{conclusion} presents suggestions for future research directions and concludes the paper.

\section{Related Work}
\label{related_work}
Coverage-Directed Test Generation (CDG) \cite{IBM2003} provided an early attempt at using artificial intelligence (AI) to automate test biasing during simulation-based verification. CDG methods, exemplified by \cite{IBM2003, Ederetal07}, use AI techniques to build a model mapping test generation parameters, to the coverage exercised by generated stimuli. The model is used to bias test generation towards plugging coverage holes to achieve faster coverage closure. 

The approach in \cite{IBM2003} utilised a Bayesian Network to learn the relationship between test generation constraints and coverage, then subsequently querying the Bayesian Network for new constraints that have a high probability of exercising a given coverage point. Although the results showed that the CDG method could indeed accelerate coverage progress, it was also clear that the method required significant domain knowledge to be encoded into the model. 

Encoding domain knowledge into an AI model is largely a manual endeavour which has to be undertaken by verification engineers who understand the design. Therefore, the problem of verification engineer time and effort being taken by manual work still remains. It could be argued that the time and effort expended during the model building phase is a one-off cost. But if the design changes significantly across projects -- a likely outcome given the ever-expanding complexity and functionality of modern hardware -- then a large part of that manual work has to be done again.

Building an AI model mapping coverage feedback to test stimuli is generally known to be difficult in practice. This can be due to several reasons, such as a lack of positive training examples; different abstraction levels between test stimuli and microarchitectural behaviour; lack of a suitable distance metric; and the lack of suitable or adequate positive training examples.

In this paper, we adapt CDG for biasing test selection, while addressing several of its shortcomings. We first solve the problem of lack of labelled training data by converting simulation trace data from an unlabelled into a labelled state. We proceed to model relationships between simulated test stimuli and coverage using supervised learning. This generalises the model building process to be performed by any classifier of choice and removes the need to extensively encode domain knowledge into the model structure. Relationships between test stimuli and coverage are learnt automatically despite different abstraction levels. We use coverage groups derived from the verification plan to approximate distance between coverage points, and to mitigate problems that would have arisen due to the lack of positive training examples when targeting coverage holes.

\section{Coverage Directed Test Selection}
\label{cds}

Coverage-directed test selection (CDS) focuses on selecting a small subset of pre-generated test stimuli based on the constraints learnt by the CDS engine. This necessitates that a large set of test stimuli have already been generated, for example through a constrained random process. 

If test stimuli generation takes a small amount of time and computation, while simulation takes an exponentially longer time plus significantly more computation, it is beneficial to generate a large number of test stimuli, then subsequently bias selection towards the most promising tests. CDS is therefore ideal for verification environments where test generation is `cheap', while simulation is `expensive'.

    \subsection{Constraint Extraction Theory}
At the core of the CDS engine is a constraint extraction process during which constraint learning and probability estimation are performed for two possible outcomes, namely exercising target coverage, versus not exercising target coverage. The constraint extraction process takes previously simulated test stimuli and coverage data as inputs.   

There are two main aspects of constraint extraction that require special attention: labelling based on coverage, and constructing training sets. We now discuss each of these in turn. 

    \subsubsection{Labelling}
Labelled data are required to train the classification model. A major issue immediately arises because the target coverage points have never been exercised before, meaning that positive training examples are not present in the data. 

To mitigate this lack of positive training examples, we first group the functional coverage space into non-overlapping coverage groups. Coverage grouping enables us to label the training set based on the coverage groups, and also build coverage feedback models at the coverage group level. We derive coverage groups through the coverage hole analysis method of partitioning  \cite{IBM2002}. Partitioning divides the coverage space into manageable partitions based on information readily obtained the verification plan. After identifying target coverage groups, each group's exercising test stimuli can be labelled as positive examples.

The CDS engine can automatically inspect a coverage report to identify target coverage groups containing coverage holes. To ensure adequate training data is obtained, only coverage groups exercised by a specified minimum number of test stimuli are considered as target coverage groups. 

\begin{figure}[t]
    \centering
    \includegraphics[width=0.49\textwidth]{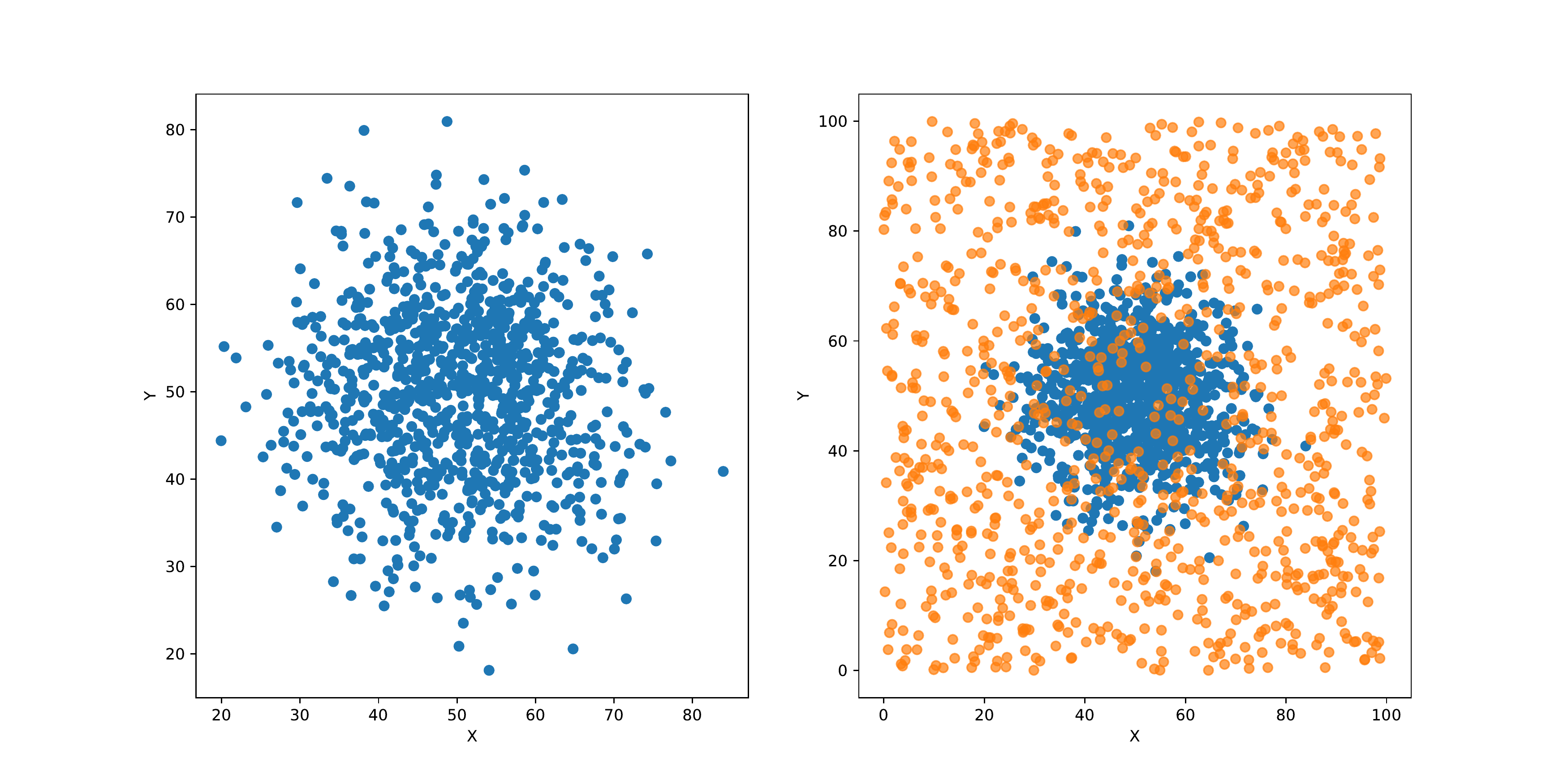}
    \caption{Original data (left). Original data augmented with reference data (right).}
    \label{fig:augmentation}
\end{figure}

    \subsubsection{Training Set Construction}
Each target coverage group's training set is constructed by adopting Hastie \emph{et al.}'s \cite{Hastieetal2005} method of transforming problems from unsupervised to supervised learning. This entails augmenting an `original' data set with `reference' data. We treat stimuli that exercised the coverage group as 'original' data generated from specific, known constraint distributions. Conversely, we draw a sample of 'reference data' of the same size from tests that did not exercise the coverage group, assuming they are generated from unknown constraint distributions. We label reference data as negative examples and pool the data sets. 

The sampling step leads to balanced training sets in which positive and negative examples are equally represented. Sampling from different constraint distributions is likely to facilitate for the positive and negative examples to be separable due to the high density of positive examples in specific regions of the feature space. 
Figure \ref{fig:augmentation} shows an example of original blue data points sampled from a normal distribution, augmented with an equal number of reference orange data points sampled from a uniform distribution. A binary classifier can then be trained to classify the original data from the reference data. It can be seen that the separation of positive and negative examples is generally made possible by data augmentation, but it is imperfect and prone to some error based on the noisiness of the data. 

    \subsection{Constraint Extraction example using a Decision Tree}

To further aid understanding, we now provide an example of the constraint extraction process using classification and regression tree (CART) \cite{Breimanetal} methodology. In general, a decision tree is a function, $f$, that takes a vector, \textbf{x}, of $n$ feature values \{$x_0, x_1, ... , x_{n-1}$\} as inputs; the function returns a decision, $\hat{y}$ (e.g., a class for classification, or a real value for regression) as the output.

Suppose that the example DUT is a small radar signal processor that either takes inputs from main memory or from a radar receiver. During processing, the inputs are written to a specific register determined by a data bin value and processed according to their size. If output is required, the processed results are stored in main memory. 

The features for training the decision tree are the DUT's configuration fields, which control the device's behaviour. The DUT's features are:
\begin{itemize}
    \item \textbf{input\_interface}, with values \{MEM, RDR\}
    \item \textbf{data\_size}, with values \{1, 2, 3, 4\}
    \item \textbf{output\_active}, with values \{0, 1\}
    \item \textbf{data\_bin}, with values \{0, ... , $2^{32}-1$\}
\end{itemize}

Broadly speaking, the multi-dimensional region defined by the configuration fields and their values forms the coverage model. This simple DUT's coverage model would consist of $6.87 $x$ 10^{10}$ cross-product coverage points if all possible combinations needed to be verified. In real designs, coverage model sizes can be reduced by, for example, largely focusing on complex functionality that is known to be bug-prone, or by excluding combinations that should never occur according the specification. 

\begin{figure}[t]
     \centering
     \begin{subfigure}[b]{0.49\textwidth}
         \centering
         \includegraphics[width=\textwidth]{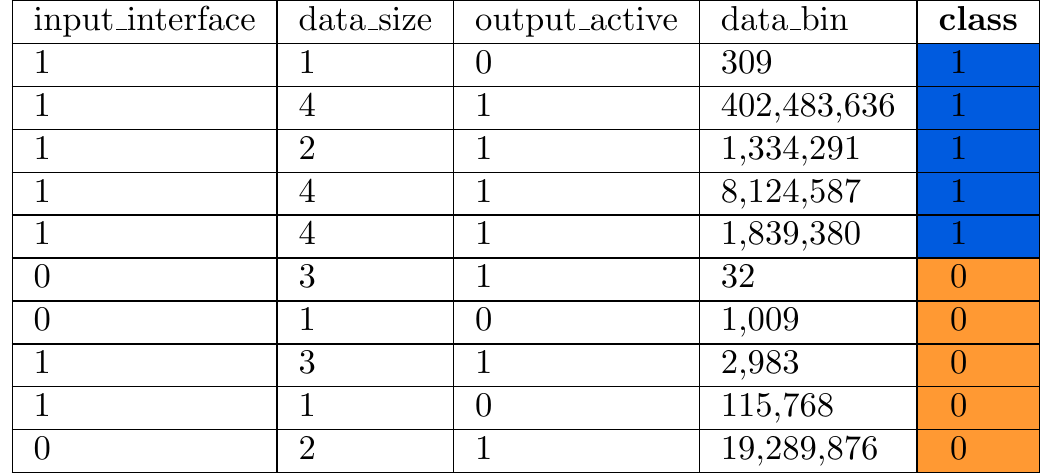}
         \caption{Training set}
         \label{fig:eg_training_set}
     \end{subfigure}
     \hfill
     \begin{subfigure}[b]{0.49\textwidth}
         \centering
         \includegraphics[width=\textwidth]{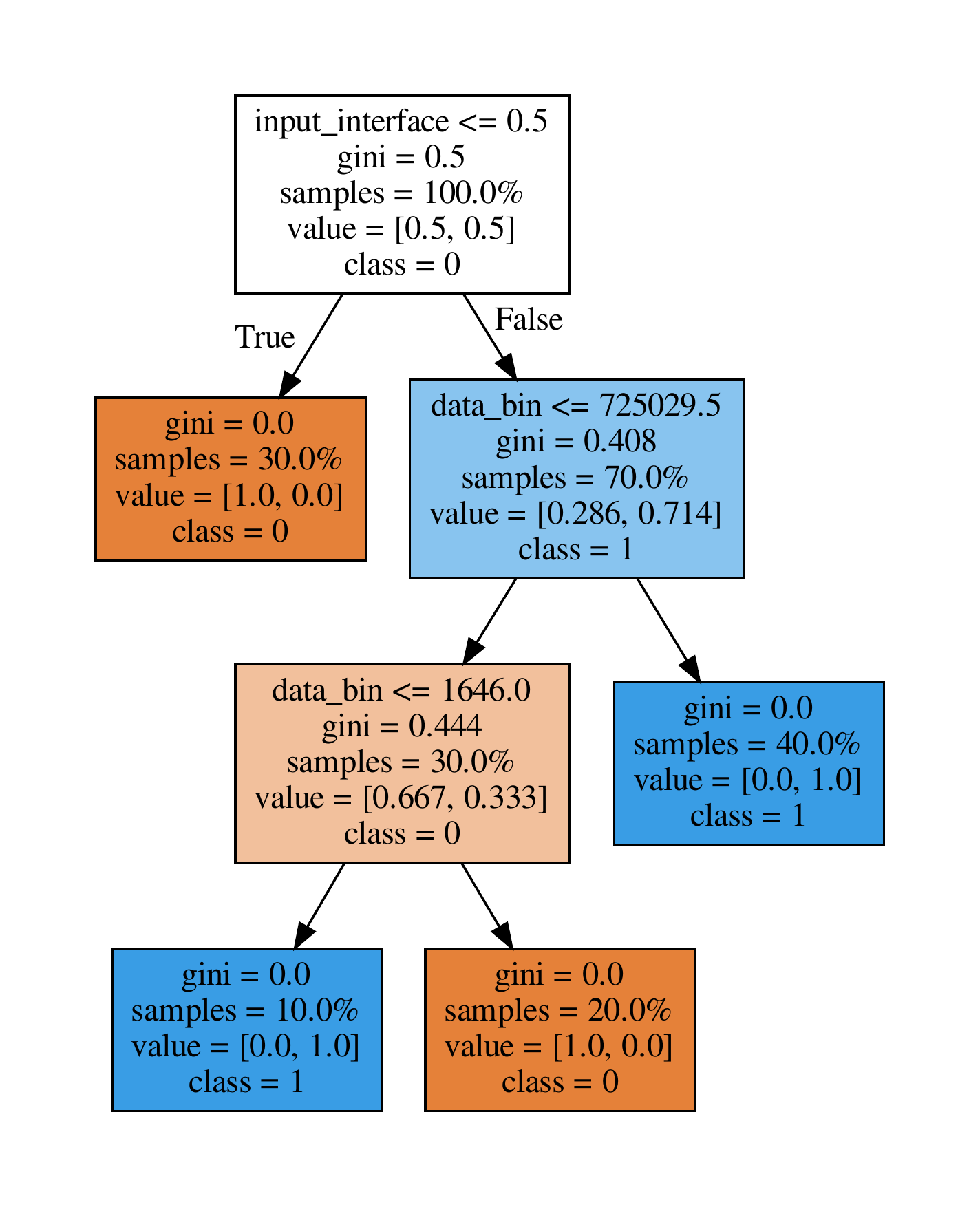}
         \caption{Decision tree}
         \label{fig:eg_dt}
     \end{subfigure}

        \caption{Decision tree induced from training set}
        \label{fig:dt_induction}
\end{figure}

A test is described as a vector of feature values, an example being (\textbf{input\_interface}: MEM, \textbf{data\_size}: 4, \textbf{output\_active}: 1, \textbf{data\_bin}: 298). Each test belongs to one of two mutually exclusive classes known \textit{apriori}, which we will denote 1 and 0 for positve and negative classes, respectively. The table in \ref{fig:eg_training_set} shows an encoded training set comprising of 10 training examples. A training set is constructed for each target coverage group. 

The target coverage group associated with the training set in Figure \ref{fig:eg_training_set} contains coverage points that are exercised when data inputs have been received from the radar receiver (\textbf{input\_interface} $=$ 1). In addition, the coverage points tend to be exercised when output is required (\textbf{output\_active} $=$ 1), as is typically the case when the value of \textbf{data\_bin} is relatively high. 

Figure \ref{fig:eg_dt} shows the decision tree induced from the example training set. The root and internal nodes display the splitting condition alongside the Gini impurity of the node (gini), the proportion of training examples remaining in the node relative to the complete training set (samples), the class frequencies of the node's training examples - one for each class (value), and the node's predicted class (class). Class frequencies are important because they can be interpreted as an empirical estimation of the model's ability to generalise to unseen data. The leaf nodes display the Gini impurity of the node, the proportion of examples used to train the node, and the node's predicted class. 

Following the decision tree, we find that the example test (\textbf{input\_interface}: 0, \textbf{data\_size}: 4, \textbf{output\_active}: 1, \textbf{data\_bin}: 298) is predicted to belong to the negative class with 100\% probability. This tells us that it is unlikely to exercise the coverage points within the target coverage group. The implication of a negative class prediction for a newly generated test is that it is not selected for simulation.

\section{Experimental Setup}
\label{experiment}

\subsection{Design Under Test}
We evaluated CDS on Infineon Technologies' Radar Signal Processing Unit (RSPU) AURIX TC3XX design. The RSPU is a large, complex and highly configurable block crucial to the function of Advanced Driver Assistance Systems (ADAS) in Infineon's AURIX family of microcontrollers. The design functions as a semi-autonomous accelerator for performing Fast Fourier Transforms (FFT) on input data. The RSPU has also been used as the DUT in research that focused on novelty-directed verification \cite{Zheng2019, Blackmoreetal2021} as a mechanism for test selection and ordering.

The behaviour of the RSPU is configured by inputs received as a set of configuration instructions and data via two input interfaces. The largest effort to uncover logical errors in the RSPU design occurs via simulation-based functional verification. 

\subsection{Test Stimuli and Coverage}
Each RSPU test is a feature vector of 300 configuration fields. Test stimuli for the DUT are generated in an industry-standard constrained random environment. The DUT is treated as a black box over which we can control the input stimuli, and observe outputs in the form of functional coverage data.

The DUT functionality we aim to verify is detailed in the verification plan as coverage groups. There are almost 200 coverage groups of varying size that collectively form the RSPU's functional coverage model. We assume that coverage points are grouped together because they are judged to be semantically similar by the verification team. 

\subsection{Metric}
The metric we use to assess the success of these experiments is functional coverage. The functional coverage model contains about 6,000 whitebox cross-product coverage points. 

\subsection{Experimentation Database}
Relying on DUT simulation was impractical because it takes a long time to obtain results. In a real-world setting, achieving 100\% functional coverage closure for the RSPU requires approximately 2 million constrained random test stimuli. It takes an average of 2 hours to simulate an RSPU test within the constrained random testbench. Achieving coverage closure would require continuous consumption of 1,000 EDA licences for 6 months, alongside several months of effort spent by verification engineers writing constraints. 

Instead of direct DUT simulation, the experiments were expedited through the creation of an experimentation database. The experimentation database stored test stimuli and coverage data that could emulate DUT simulation. A golden regression of approximately 3,000 tests were found to enable optimal coverage closure on the RSPU's coverage model. These golden regression tests were stored in the experimentation database to ensure coverage closure was possible. 

To resemble real-life RSPU verification, where there are relatively more tests that do not contribute to coverage closure, a further 83,000 constrained random test stimuli were added to the database. Because they are constrained random tests, they are also able to exercise coverage on the DUT. However, the coverage they exercise will generally be easy to reach, hence we cannot reasonably expect them to exercise new coverage, particularly during the advanced stages of verification when mostly hard-to-reach coverage points remain. These extra 83,000 stimuli were estimated to be large enough in number to make the selection problem more realistic by introducing diversity and redundancy among the tests, but small enough to enable most experiments to be completed within a day. 

Before experimentation began, the 86,000 tests in the database were simulated, and the coverage achieved by each test was saved to a separate table in the database. Relations were defined between the test and coverage tables to associate each test to the coverage points it activated. During experimentation, test selection means sampling without replacement from the 86,000 stimuli. Simulating a test means querying the database for the coverage points exercised by the test. The experimentation database allows for tests to be simulated in any order, and to quickly query for the coverage points relating to a particular test, without direct DUT simulation. This enabled rapid experimentation and prototyping of solutions.

\subsection{CDS Classifiers}
To understand the trade-offs associated with the different classifiers, we distinguish between tree-based and non-tree-based classification. Tree-based classifiers depend on decision trees as the mechanism for making predictions, while non-tree-based classifiers depend on other mechanisms for their predictions. CDS experiments were implemented using Python3. All classifier implementations used in the experiments are available from Scikit-learn \cite{sklearn2022}. 

The following decision tree-based classifiers were compared:
\begin{itemize}
    \item Baseline - a `dummy' classifier that makes predictions in a uniform random manner regardless of the input features. A dummy classifier is useful for gauging baseline performance when comparing multiple classifiers.
    \item DT - decision tree classifier with standard hyperparameters.
    \item DCDT - depth-constrained decision tree whose maximum depth was chosen based on the observation that many cross-product coverage points in the coverage model tend to encompass 3 configuration fields. Therefore, the maximum depth was set to 3. This is an experiment to determine the effect of integrating domain knowledge into the model. One way of integrating domain knowledge into a machine learning model is through hyperparameter tuning.
    \item DCRDT - depth-constrained decision tree with randomised splits. Similar to DCDT, the maximum depth for the induced trees is restricted to 3. But instead of using an information gain measure to determine which feature to split on, the algorithm randomly selects a feature to split on from the set of available features.
    \item RF - random forest algorithm consisting of several decision trees. 
    \item GB - gradient boosting algorithm based on decision tree estimators.
\end{itemize}

The following non-tree classifiers were also compared: 
\begin{itemize}
    \item LR - logistic regression classifier.
    \item NN - five-layer neural network.
    \item NB - Naive Bayes classifier.
\end{itemize}

\subsection{CDS Procedure}
CDS is performed as part of an iterative simulation-based verification process. In the early stages of verification, 1,000 tests are randomly selected and simulated per iteration. When 90\% functional coverage is reached, which typically takes roughly 5,000 tests, CDS is activated. 

During each CDS iteration, a classification model is trained for each target coverage group. The trained model is used to select the most optimal test to simulate in the next iteration, such that the total number of tests simulated in each iteration is equal to the number of target coverage groups. Since it typically takes less than 5 minutes to train the models, we fully re-train the models during every iteration. 

No feature selection is performed for the CDS experiments, and only basic feature engineering is performed. To avoid increasing the number of features, an ordinal encoding scheme is applied to the categorical features. Features that can take on a large range of values are categorised according to powers of 2, thereby reducing the cardinality of the features and facilitating better model generalisation.

\section{Results}
\label{results}

We evaluate the performance of the underlying CDS classifier based on the reduction in the number of tests required to reach certain functional coverage levels. The more effective the underlying CDS classifier is at learning from coverage feedback, the more effective it is at selecting test stimuli that target coverage holes, thereby resulting in fewer tests being simulated.  

The number of tests required to reach 95\%, 98\% and 99\% functional coverage were recorded. Performing CDS after achieving 99\% functional coverage was not significantly better than constrained random testing for this particular coverage model. Because roughly 60 coverage points remain unexercised at that point, we reasonably expect that a verification engineer can intervene and manually bias testing towards exercising them. 

\begin{table*}[t]
\centering
\begin{tabular}{|c ||c |c |c |c |c |c |c |} 
 \hline
 \textbf{Functional Coverage} & \textbf{Random} & \textbf{Baseline}& \textbf{DT} & \textbf{DCDT} & \textbf{DCRDT} & \textbf{RF} & \textbf{GB} \\ [0.5ex] 
 \hline
 95\% & 12866 & 11287 & 12561 & 11770 & 10659 & 12329 & 10911 \\
 \hline
 98\% & 29300 & 30374 & 29396 & 27179 & 26553 & 28440 & 25801 \\
 \hline
 99\% & 44200 & 44582 & 42334 & 41458 & 40628 & 44044 & 39419 \\
 \hline
 \textbf{Savings (vs. Random)} & & & & & & & \\
 \hline
 95\% &  & -12.27\% & -2.37\% & -8.52\% & -17.15\% & -4.17\% & -15.2\% \\
 \hline
 98\% &  & +3.67\% & +0.33\% & -7.24\% & -9.38\% & -2.94\% & -11.94\% \\
 \hline
 99\% & & +0.86\% & -4.22\% & -6.2\% & -8.08\% & -0.35\% & -10.82\% \\ [1ex] 
 \hline
\end{tabular}
\caption{Number of tests required to reach given functional coverage levels: Decision Tree CDS techniques vs Baseline and Random}
\label{savings_tbl_cds_initial}
\end{table*}

\begin{table}[h!]
\centering
\begin{tabular}{|c ||c |c |c |c |} 
 \hline
 \textbf{Functional Coverage} & \textbf{Random} & \textbf{LR} & \textbf{NN} & \textbf{NB} \\ [0.5ex] 
 \hline
 95\% & 12866 & 10420 & 10065 & 9614 \\
 \hline
 98\% & 29300 & 27282 & 24786 & 22731 \\
 \hline
 99\% & 44200 & 42667 & 37919 & 35960 \\
 \hline
 \textbf{Savings (vs. Random)} & & & & \\
 \hline
 95\% &  & -19.01\% & -21.77\% & -25.28\% \\
 \hline
 98\% &  & -6.89\% & -15.41\% & -22.42\% \\
 \hline
 99\% & & -3.47\% & -14.21\% & -18.64\% \\ [1ex] 
 \hline
\end{tabular}
\caption{Number of tests required to reach given functional coverage levels: non-tree-based CDS techniques vs Random}
\label{savings_tbl_nts_nontree_based}
\end{table}

\subsection{CDS using Tree-Based Classifiers}
We begin by studying the performance of CDS when decision tree-based classifiers are utilised for constraint extraction. We are interested in these because the resulting constraints can be interpreted by human beings when required. Their ability to reduce simulation resource consumption is compared to random selection, which is analogous to traditional constrained random testing. 

The results from decision tree-based CDS experiments are summarised in Table \ref{savings_tbl_cds_initial}. The most important measures of CDS classifier performance in the table are the simulations saved at 99\% functional coverage. 

The dummy classifier had the worst performance by actually adding to the average number of simulations performed by random selection. This is to be expected because the dummy classifier is conceptually equivalent to random selection. 

The standard decision tree algorithm's performance was inconsistent, as it can be seen to add more simulation at 98\% coverage, yet at the same time saving simulation at 95\% and 99\% coverage. This could be due to a well-known issue with untuned decision trees: they tend to overfit to the training data, making predictions that generalise poorly to unseen data. This issue is clearly mitigated when the depth of the decision tree is constrained (DCDT and DCRDT classifiers), which is a known method of combatting overfitting. 

The best performer among decision tree-based algorithms is the gradient boosting classifier, reducing the number of stimuli required to reach 99\% functional coverage by almost 11\% and consistently performing better than random selection at every functional coverage level. 

\subsection{CDS using Non-Tree-Based Classifiers}
For completeness, we also studied the performance of CDS using a selection of classifiers that are not based on decision trees. The results are summarised in Table \ref{savings_tbl_nts_nontree_based}. 

The worst performing non-tree-based classifier is logistic regression, which utilises a linear estimator that is unlikely to capture the nonlinear complexities in the coverage data. It is reasonable to expect that the logistic regression algorithm would be the worst performing out of the three. 

The best performing classifier is Naive Bayes, managing to save 18.64\% of tests being simulated when compared to random selection. However, this performance comes at the expense of constraints that are not transparent to human beings. 

Non-tree-based classifiers generally achieved higher savings than tree-based classifiers. However, non-tree-based algorithms also tend to be slower to train, and demand more computational resources. As an example, the neural network (NN) took 5 days to complete the CDS experiment, whereas the standard decision tree (DT) completed the experiment in 1 day. However, in practice, the extra time required by the neural network might be justifiable because it is more accurate than the decision tree. We can check this with some quick calculations. The neural network achieved 99\% functional coverage with 4,415 less tests than the decision tree. For the RSPU, this translates to approximately 9,000 hours of saved simulation time. Assuming a maximum simulation capacity of 1,000 tests a day, the neural network saves 9 days' worth of simulation resources compared to the decision tree. Accounting for the extra 4 days of training and inference time required by the neural network still leaves a net simulation resource saving of 5 days. If the prime objective is to reduce simulation, a neural network classifier would be preferable to a decision tree in this case.

\section{Conclusion}
\label{conclusion}

In this work, we have proposed coverage-directed test selection as a method for automatic test biasing during simulation-based verification. Coverage-directed test selection uses machine learning on coverage feedback to identify and prioritise test stimuli that target coverage holes. Coverage-directed test selection adds value to simulation-based verification through gains in testing efficacy (targeted stimuli) and testing efficiency (reduced simulation). These enhancements lead to faster functional coverage closure, higher quality test stimuli and reduced manual effort throughout the simulation-based verification process. 

According to the experimental results, this particular RSPU coverage model could be nearly closed by simulating 18\% less tests when using coverage-directed test selection. Scaling to RSPU verification in production, this could mean simulating 1.7 million, instead of 2 million tests, to achieve coverage closure. Note that the golden regression would still consist of 3,000 tests, hence there is still significant redundancy in the coverage-directed test selection. We are working to improve feature selection, feature engineering, and dimensionality reduction during the constraint extraction process to improve these results. 

Although coverage is a useful measure of verification progress, a functional coverage model is a somewhat subjective approximation of the design behaviour that must be explored. Because of the subjectivity inherent in the coverage model, and because bugs can hide anywhere in the design, there is value in aiming to explore beyond the coverage model. 

Future research will focus on further optimising test selection by combining coverage-directed test selection method with novelty-driven test selection \cite{Blackmoreetal2021}. In contrast to coverage-directed test selection, novelty-driven test selection has been found to be useful at exploring more of the design by prioritising relatively novel tests for simulation. We expect that complementing coverage-directed test selection with novelty-driven test selection will result in improvements in the test selection capabilities of both methods.

\bibliographystyle{IEEEtran}
\bibliography{main}

\end{document}